\documentclass[a4paper,10pt,twoside,slidetop,cjk]{cpc-hepnp}

\usepackage{multicol}
\usepackage{graphicx}
\usepackage{booktabs}
\usepackage{amssymb,bm,mathrsfs,bbm,amscd}
\usepackage[tbtags]{amsmath}
\usepackage{lastpage}
\usepackage{color}
\usepackage{leftidx}

\begin{document}

\fancyhead[c]{\small Submitted to Chinese Physics C}
\fancyfoot[C]{\small 010201-\thepage}

\footnotetext[0]{Received 31 June 2015}

\title{IsoDAR Neutrino Experiment Simulation with Proton and Deuteron Beams
\thanks{This work is partly supported by the national 973 program (2014CB845406), the National Natural Science Foundations of
China under the Grants Number (11175220) and Century Program of
Chinese Academy of Sciences (Y101020BR0). } }

\author{%
     Fengyi Zhao$^{1}$%
\quad Yao Li$^{1,2}$%
\quad Chengdong Han$^{1£¬2}$
\quad Qiang Fu$^{1,2}$%
\quad Xurong Chen$^{1)}$\email{xchen@impcas.ac.cn}%
}
\maketitle

\address{%
$^1$ Institute of Modern Physics, Chinese Academy of Sciences, Lanzhou 730000, China\\
$^2$ University of Chinese Academy of Sciences, Beijing 100049, China\\
}

\begin{abstract}

In this paper we consider high-intensity source of electron antineutrinos from the production and subsequent decay of 8Li.
It opens a wide range of possible searches
for beyond standard model physics via studies of the inverse beta decay interaction.
In IsoDAR experiments Lithium 8 is a short lived beta emitter producing a high
intensity anti-neutrinos, which is very suitable for making several important neutrino experiments.
In this paper we used the GEANT4 program. to simulate neutrino production using proton and deuteron beams. We find that the
neutrino production rate is about 3 times from deuteron beam than from proton beam
in low energy region.

\end{abstract}



\footnotetext[0]{\hspace*{-3mm}\raisebox{0.3ex}{$\scriptstyle\copyright$}2013
Chinese Physical Society and the Institute of High Energy Physics
of the Chinese Academy of Sciences and the Institute
of Modern Physics of the Chinese Academy of Sciences and IOP Publishing Ltd}%

\begin{multicols}{2}

\section{Introduction}

As the next step, neutrino physics requires high intensity experiments for
precision measurements.
We propose the neutrino experiment on proton or deuteron beam hits target.
The Isotope Decay At Rest (IsoDAR)~\cite{adelmann}~\cite{isodar} technique provides a high-intensity, low energy
source of antineutrinos with sensitivity to antineutrino
oscillations. The experiment can
perform compelling tests of models for new physics that
explain high $\Delta$m$^2$ oscillations through the introduction of
one or more sterile neutrinos.

IsoDAR can emit anti-electron neutrino at a rather
high-flux by beta decay of short lived isotope. Lithium 8 is a nice candidate of
this isotope and it can be produced by injection of proton or deuteron on
a beryllium 9 target. To meet the required neutrino flux of oscillation experiment,
the intensity of proton beam should be at least 10 mA with energy of 60 $\sim$ 1000 MeV.
This requirement is the main limitation to IsoDAR source for the last decades.
But now, by taking advantage of high intensity proton beam technique,
this becomes very promising and worth a detailed simulation.
IsoDAR sources provide an opportunity for the precision
measurements required for the future of neutrino physics~\cite{conrad2014}.

\section{IsoDAR Design}

\begin{center}
\includegraphics[width=9cm]{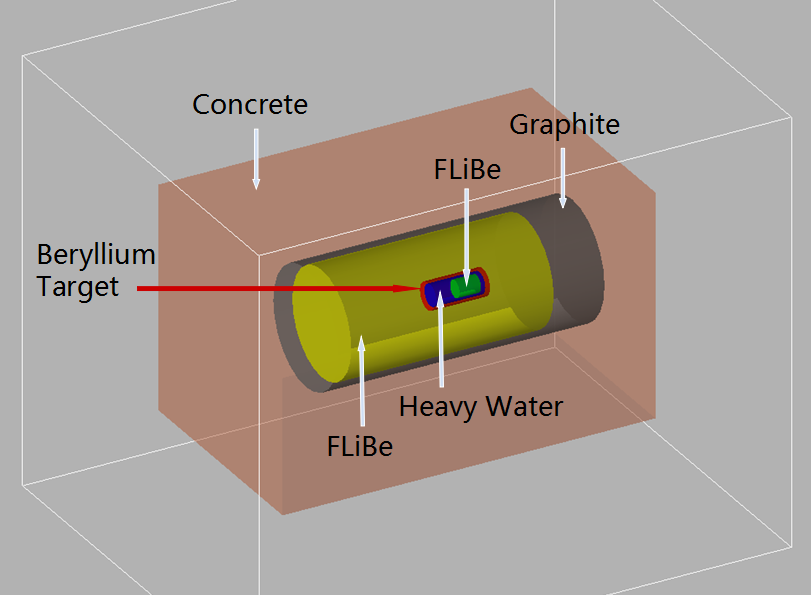}
\figcaption{\label{fig1} Initial design for IsoDAR target. }
\end{center}

Fig.~\ref{fig1} shows initial design for IsoDAR target (Ref.~\citep{lab1}). It's like
a neutrino converter: absorb proton (or deuteron) and then emit neutrino.

 When proton inject on the Beryllium target, several interactions happen subsequently:
\begin{eqnarray}
\label{eq1}
p+\leftidx{^9}Be\longrightarrow\leftidx{^8}Li+2p
\end{eqnarray}

\begin{eqnarray}
\label{eq2}
p+\leftidx{^9}Be\longrightarrow n+p+2\ \leftidx{^4}He
\end{eqnarray}

\begin{eqnarray}
\label{eq3}
n+\leftidx{^7}Li\longrightarrow\leftidx{^8}Li
\end{eqnarray}

\begin{eqnarray}
\label{eq4}
\leftidx{^8}Li\longrightarrow\overline{v}_e+e+\leftidx{^8}Be
\end{eqnarray}

With proton beam incidences, the Beryllium 9 target can produce Lithium 8 and neutron,
to which we pay more attention, as Eq.~\ref{eq1} and Eq.~\ref{eq2} describes.
Also, we can see from Eq.~\ref{eq4} that the production of neutrino is directly related
to that of Lithium 8. So in our primary design, there is a FLiBe sleeve outside and inside
of the Be target respectively to absorb neutrons and produce Lithium 8, which process is
described by Eq.~\ref{eq3}. FLiBe is molten salt made from a mixture of lithium
fluoride (LiF) and beryllium fluoride (BeF$_{2}$) with abundance of Li$_7$ up to 99.99\%.
It has a high melting point and a better mechanical property compare to elemental Lithium.
The heave water between Be target and inner FLiBe sleeve is moderator for neutron to improve
its absorptivity by Li$_7$, besides, the heavy water sleeve can be design as cooling system
to carry out heat from beam injection. The graphite layer outside the outer FLiBe sleeve is
set as a reflector to improve the utilization for neutron. To prevent neutron radiation,
a thick concrete sleeve is set outside covering the whole equipment as a shielding layer.

While using Geant4~\cite{geant4} to simulate interactions of neutrino production, we need to specify
relevant model for every process and every particle, i.e., the PhysicsList package in Geant4.
In our primary simulation, we choose QGSP\_BIC\_HP, a reference physics list from Geant4,
to describe relevant reaction channels and cross section. This package implies quark gluon
string model to high energy interaction (5-25 GeV), the excited nucleus created is
then passed to the pre-computed model modelling the nuclear de-excitation.
As for interactions below 10 GeV, it uses binary cascade model (BIC) to
describe the propagation of incident hadron and secondary through nucleus,
which is regard as a series of two-particle collision. For energy from
20 MeV down to thermal particle, this model takes data driven high precision
neutron package (NeutronHP) to transport neutrons below 20 MeV down to thermal energy.
Compare to other physics list package, QGSP\_BIC\_HP is able to rebuild the production
of secondary particles from $p-A$ or $n-A$ interactions well and is more precise
in low energy region. It can gives us a more accurate isotope production rate
and neutrino distribution in IsoDAR.

\section{Simulation Results}

In the present work, the GEANT4 simulation code [3]
has been used to simulate the low-energy protons induced
isotopes production in Be targets.

After a simulation of a number of proton incidence, we can get the neutron
spectra crossing different barrier components as Fig.~\ref{fig2} shows:
\begin{center}
\includegraphics[width=9cm]{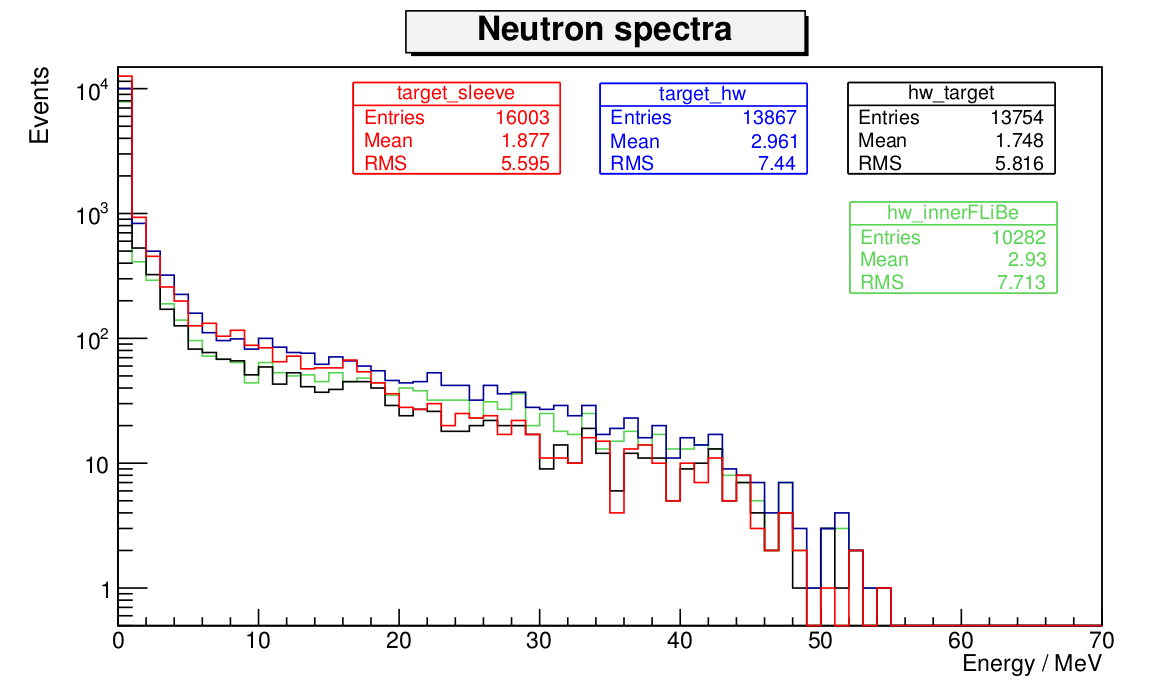}
\figcaption{\label{fig2} Spectra for neutrons crossing the barrier components when $10^5$ protons incident with energy of 60 MeV. }
\end{center}
From the above figure we can see the following points:

\begin{enumerate}
  \item A mass of neutrons are produced after protons incidence. And the neutrons coming from Be target to heavy water sleeve has the highest average energy.
  \item We can see the moderator effect of heavy water.
  \item Many neutrons escape out from the target sleeve, so it's necessary to add a
        thick FLiBe sleeve surrounding target to have a good usage of neutrons.

\end{enumerate}

Besides, we can get the isotope production in different parts from proton incidence and deuteron incidence.
\begin{center}
\includegraphics[width=7cm]{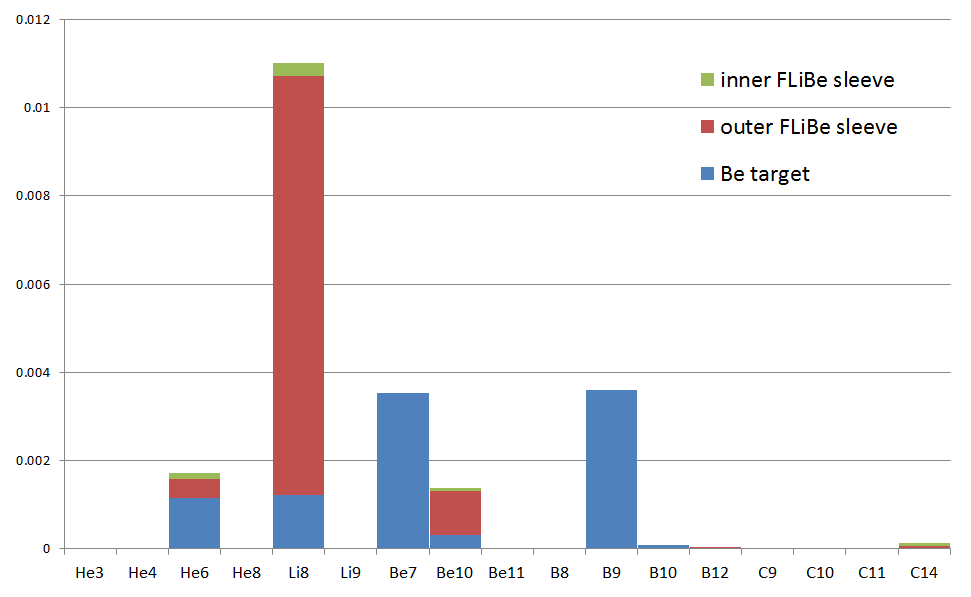}
\\
\includegraphics[width=7cm]{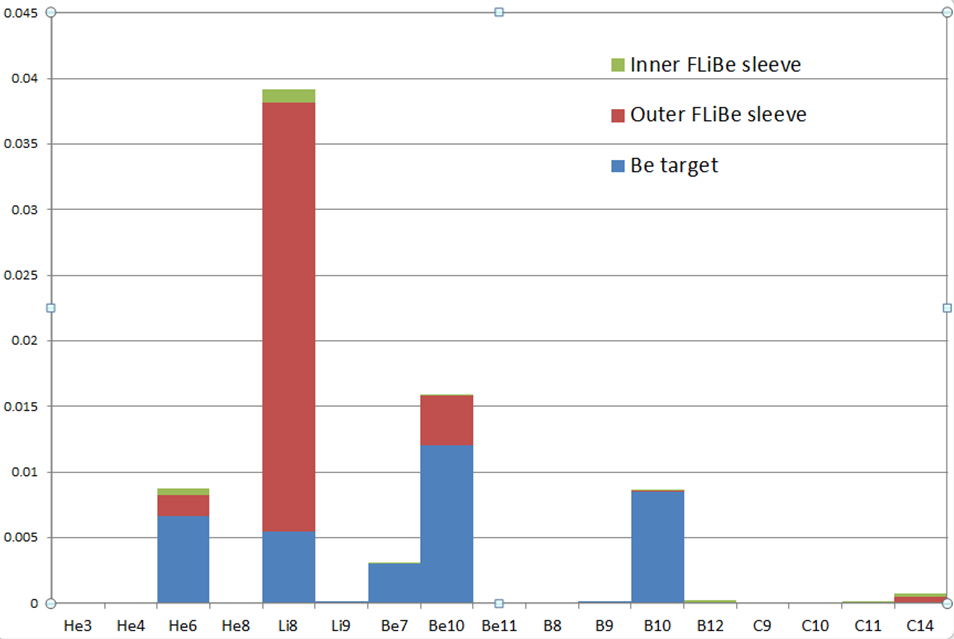}
\figcaption{\label{fig3}Isotope production in different parts. Top: 60 MeV proton incidence; Bottom: 80 MeV deuteron incidence. }
\end{center}
We can see from Fig.~\ref{fig3} that Li8 is mainly produced in the outer FLiBe sleeve,
which is reasonable as more neutrons enter into this sleeve. Also, we can see that
Deuteron's Li8 production rate is higher than Proton's. This result is meaningful
and worthy of further research as it will tell us which incident beam is better
in IsoDAR experiment. So we have a comprehensive comparison between proton and deuteron incidence with different energies.

\begin{center}
\includegraphics[width=8cm]{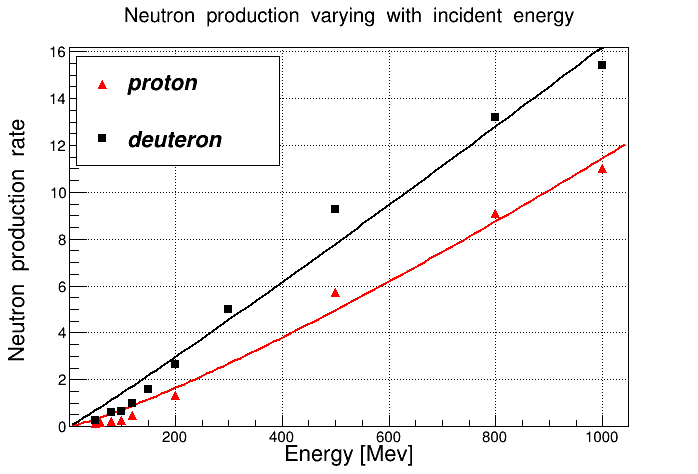}
\\
\includegraphics[width=8cm]{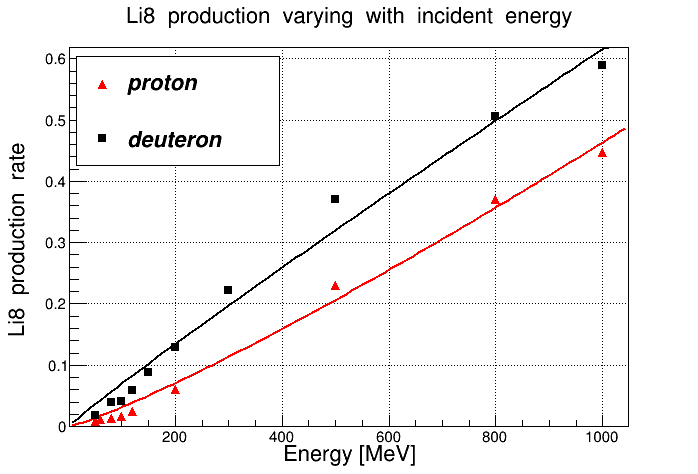}
\figcaption{\label{fig4} Neutron (Up) and Li8 (bottom) production rate at different incident energies, for proton and deuteron beams respectively. }
\end{center}

Fig.~\ref{fig4} shows the production rate of Neutron and Li8 respectively
with different incident energy for proton and deuteron beams.
we can see approximately a linear relation between production rates and incident energies,
We fit the two plots and get the following formulas.
For the neutron production rate,

\begin{eqnarray}
\label{eq5}
N_{Pn} = 2.7 \times 10^{-3} \times E^{1.2},
\end{eqnarray}

\begin{eqnarray}
\label{eq6}
N_{Dn} = 1.1 \times 10^{-2} \times E^{1.1}.
\end{eqnarray}

and for Li8,\\
\begin{eqnarray}
\label{eq7}
N_{PLi8} = 1.4 \times 10^{-4} E^{1.2},
\end{eqnarray}

\begin{eqnarray}
\label{eq8}
N_{DLi8} = 9.1 \times 10^{-4} \times E^{0.94}.
\end{eqnarray}
Here, $N_{Pn}$ and $N_{Dn}$ corresponds to the neutron production rate for proton incidence
and deuteron incidence respectively, while $N_{PLi8}$ and $N_{DLi8}$ means the Li8 production rate.
$R^2$s in the above plots are close to 1, which means this linear fitting is pretty accurate.
Compare proton's production rate at 100 MeV and deuteron's production rate at 200 MeV, we can
see that the later is not simply twice as many as the former. This indicates that our simulation
includes the nucleus effect, which is more realistic.

To have an intuitive comparison between proton and deuteron beams, we have this picture below.
\begin{center}
\includegraphics[width=9cm]{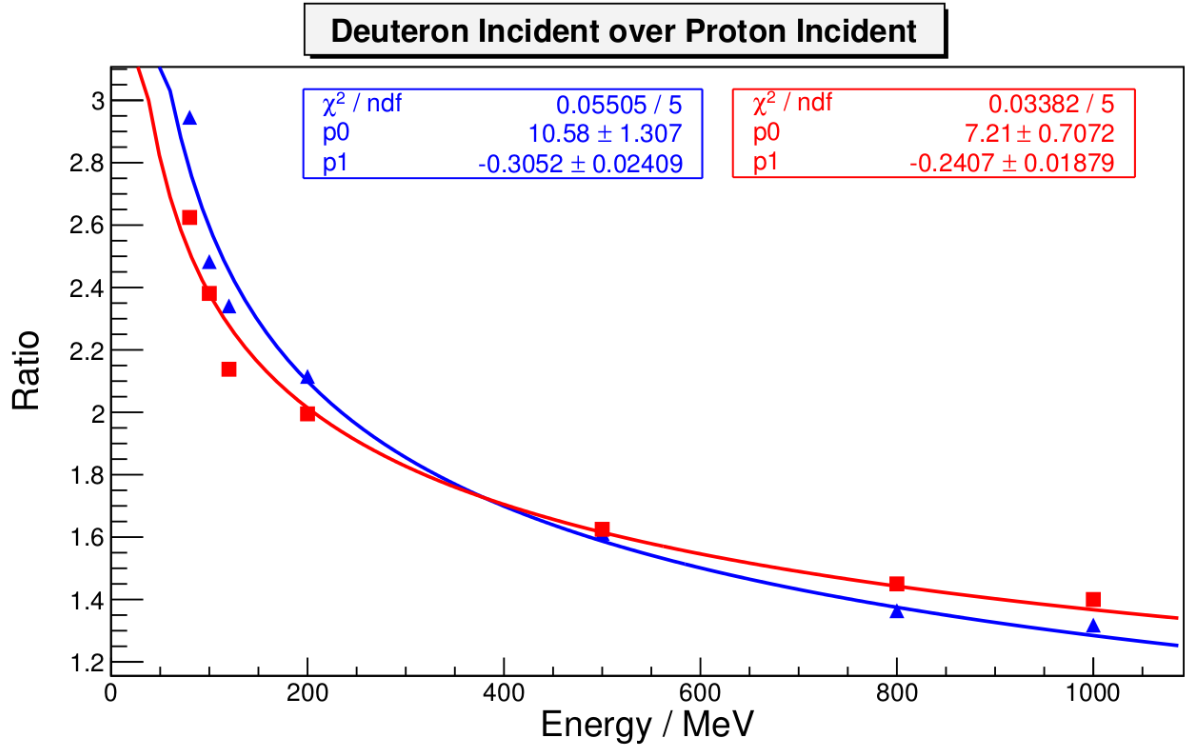}
\figcaption{\label{fig5} Li8 production ratio (Blue) and neutron production ratio (Red) of
Deuteron injection over Proton injection. This two sets of data have been fitted with a power function: ratio = p$_0 \times E^{p_1}$}
\end{center}

In Fig.~\ref{fig5} we compare the neutron and Li8 production rate for proton and deuteron
incidence with different energy, and then fitted them with power function respectively.
We can see that in the low energy region around 100 MeV, the Li8 production rate
for deuteron is about three times of that for proton. But as beam energy increases,
the ratio approaches to about 1.5. So we can expecet that several hundred MeV deuteron beam has much advantages than
proton beam with half energy to generate neutrino.

\begin{center}
\includegraphics[width=8cm]{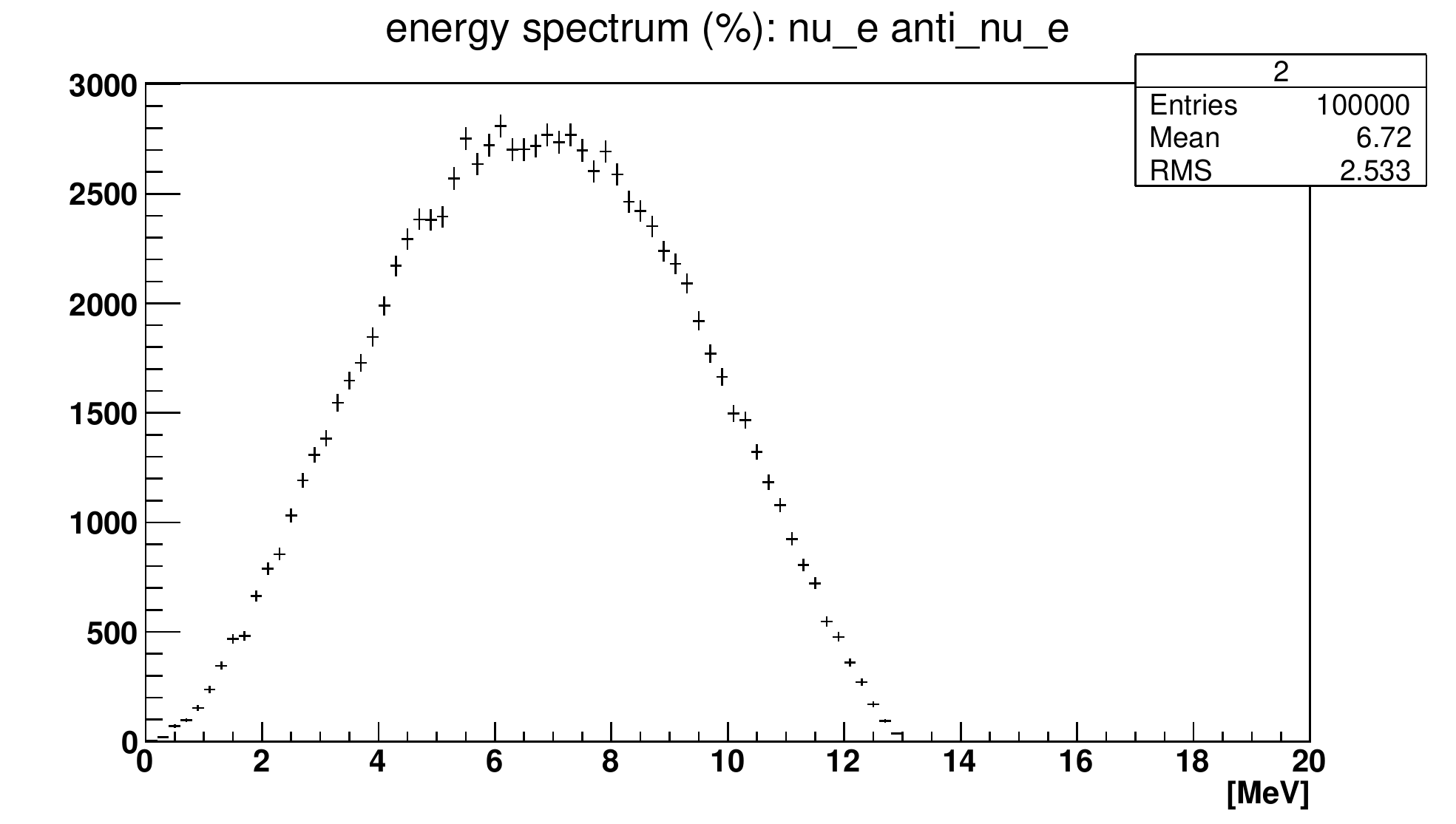}
\figcaption{\label{fig6} The 8Li isotope DAR anti-electron-neutrino flux spectrum distribution.}
\end{center}

We simulate one hundred thousand Li8 decay at rest.
The neutrino spectrum produced from IsoDAR Li8 is shown in Fig.~\ref{fig6}.
we can see that the spectrum is in agreement with experimental result~\cite{Adelmann2014}.

\section{Discussions and Conclusions}

IsoDAR represents both a novel concept of application to neutrino physics
measurements. Neutrino experiment on the China Accelerator Driven
Sub-critical System (ADS) is being proposed~\cite{zhan2015}.
The first phase of ADS is named China Initial ADS (CIADS). The CIADS facility
will offer the highest intensity protons or deuteron beams over the world
with energy between dozens of MeV and 1 GeV.

IsoDAR is a novel design, extensive and precise simulations targeting the most challenging
aspects of experiments are highly required.
In this preliminary simulation, we can see that by taking advantage of ADS technique, IsoDAR is a
feasible neutrino source for experiment searching for sterile neutrino. And we recommend deuteron beam
to hit the target as it has advantages over proton below 1 GeV.
As we see, the deuteron beam has much advantage over proton beam around several hundred MeV.
It will offer an opportunities for high precision neutrino measurements, such as
sterile Neutrino Searches, precision electroweak tests of the standard model, and coherent neutrino scattering, etc.

\end{multicols}

\vspace{-1mm}
\centerline{\rule{80mm}{0.1pt}}
\vspace{2mm}
\begin{multicols}{2}

\end{multicols}

\clearpage


\begin{thebibliography}{90}

\vspace{3mm}

\bibitem{adelmann}A. Adelmann, et al, Cost-effective Design Options for IsoDAR,  arXiv:1210.4454.

\bibitem{isodar} R. Alba, M. Barbagallo et al. arXiv:1208.1713v1 [nucl-ex], 2012.

\bibitem{conrad2014} J. M. Conrad et al, Precision $\bar ¦Í_e$-electron Scattering Measurements with IsoDAR to Search for New Physics, Phys Rev. D 89, 072010 (2014).

\bibitem{lab1} A. Bungau, R. Barlow et al. Target Studies for The Production of Lithium8 for
Neutrino Physics Using a Low Energy Cyclotron. Proceedings of IPAC2012, New Orleans, Louisiana, USA, 2012. 4145-4147

\bibitem{lab2} R. Alba, M. Barbagallo et al, Measurement of neutron yield by 62 MeV proton beam on a thick Beryllium target, Annals of Nuclear Energy 62 (2013) 590¨C595.

\bibitem{geant4} GEANT4 - a toolkit for the simulation of the passage of particles through matter, version geant4.9.5.p01: http://geant4.web.cern.ch/geant4/.

\bibitem{Adelmann2014} A. Adelmann, et al, Cyclotrons as Drivers for Precision Neutrino Measurements, Volume 2014, Article ID 347097, Advances in High Energy Physics.

\bibitem{zhan2015} Wenlong Zhan, 11th Peng Huanwu theoretical physics forum, Lanzhou, May 16, 2015.

\end{thebibliography}
\end{document}